# TENET: A TIME-REVERSAL ENHANCEMENT NETWORK FOR NOISE-ROBUST ASR

*Fu-An Chao[1], Shao-Wei Fan Jiang[1], Bi-Cheng Yan[1], Jeih-weih Hung[2], Berlin Chen[1]*

[1]National Taiwan Normal University, Taipei, Taiwan
[2]National Chi Nan University, Nantou, Taiwan

{fuann, swfanjiang, bicheng, berlin}@ntnu.edu.tw, jwhung@ncnu.edu.tw

## ABSTRACT

Due to the unprecedented breakthroughs brought about by deep learning, speech enhancement (SE) techniques have been developed rapidly and play an important role prior to acoustic modeling to mitigate noise effects on speech. To increase the perceptual quality of speech, current state-of-the-art in the SE field adopts adversarial training by connecting an objective metric to the discriminator. However, there is no guarantee that optimizing the perceptual quality of speech will necessarily lead to improved automatic speech recognition (ASR) performance. In this study, we present TENET[†][*], a novel Time-reversal Enhancement NETwork, which leverages the transformation of an input noisy signal itself, i.e., the time-reversed version, in conjunction with the siamese network and complex dual-path transformer to promote SE performance for noise-robust ASR. Extensive experiments conducted on the Voicebank-DEMAND dataset show that TENET can achieve state-of-the-art results compared to a few top-of-the-line methods in terms of both SE and ASR evaluation metrics. To demonstrate the model generalization ability, we further evaluate TENET on the test set of scenarios contaminated with unseen noise, and the results also confirm the superiority of this promising method.

## 1. INTRODUCTION

Nowadays, with the remarkable advances in deep learning and deep neural networks (DNN), a wide variety of speech-related applications have sprung up to support human activities in daily life. However, when deploying these applications into real-world scenarios, the respective performance might be seriously degraded because of environmental interferences such as noise, reverberation, background speakers, and the like. To mitigate these deteriorating effects, researchers and practitioners have designed and developed a great number of techniques. Among them, speech enhancement (SE) has seen widespread adoption as a pre-processing stage to be a crucial component connecting downstream applications such as automatic speech recognition (ASR) to alleviate environmental interferences.

In recent years, SE techniques have benefited from the impressive success of deep learning, boosting a huge body of research to explore DNN-driven methods, and these methods can be broadly categorized as either mapping-based or masking-based. For example, Lu *et al*. [1] first developed a mapping-based SE with a deep denoising autoencoder (DDAE) to map the power spectrum of noisy speech to that of its clean counterpart directly. On the other hand, Wang *et al*. [2] first employed a DNN-based model to implicitly predict a time-frequency (T-F) mask applied to a noisy spectrogram as a feasible masking-based SE method. Notably, how to pursue an effective mask [3][4][5] for speech feature representations has become one of the most prominent research topics in the recent decade.

Since the aim of SE is to improve the quality and/or intelligibility of speech, a variety of sophisticated methods are well-studied to improve the human auditory perception or optimize the corresponding objective metrics like perceptual evaluation of speech quality (PESQ) [6] and short-time objective intelligibility (STOI) [7]. For instance, deep feature loss (DFL) [8] adopted a pretrained audio scene classification network to minimize the distance between clean speech and its enhanced counterpart in latent spaces, which not only gained considerable improvements on the quality metrics but also the perceptual evaluation of human listeners. Going one step further, phone-fortified perceptual loss (PFPL) [9] was proposed along this direction by using a pretrained *wav2vec* [10] model. In addition to minimizing the above perceptual losses for SE, another choice is to optimize the target metrics directly by utilizing the generative adversarial network (GAN), such as HiFi-GAN [11] and MetricGAN [12]. Note also that MetricGAN has demonstrated to achieve superior SE performance by connecting the PESQ metrics to the discriminator.

As for the state-of-the-art SE methods, systems that optimize the perceptual quality usually can yield significant improvements in terms of SE metrics. However, due to the discrepancy of the training objectives between SE and ASR, these improvements do not necessarily translate well into downstream ASR performance, especially when an ASR system is trained with a huge amount of speech data contaminated by various types of noise. To tackle this problem, we propose a novel Time-reversal Enhancement NETwork, termed TENET, which leverages the transformation of an input noisy signal itself to form a robust SE framework for noise-robust ASR. One of the particularities of TENET is to employ both the original and time-revered versions of speech signals in the training set to learn the SE model. According to the evaluation results, this novel framework can push the limits of any kind of SE models to improve their denoising performance. Compared to some top-of-the-line methods, the best of our configurations that equips a masking-based SE module can achieve current state-of-the-art results on the VoiceBank-DEMAND dataset with respect to both SE and ASR evaluation metrics. Furthermore, the subsequent experiments also confirm the practical applicability of TENET in terms of the model flexibility and generalization ability.

---

[†] Inspired by the movie – TENET, Christopher Nolan, 2020.

[*] Some of the enhanced audio samples can be found from https://fuann.github.io/TENET.

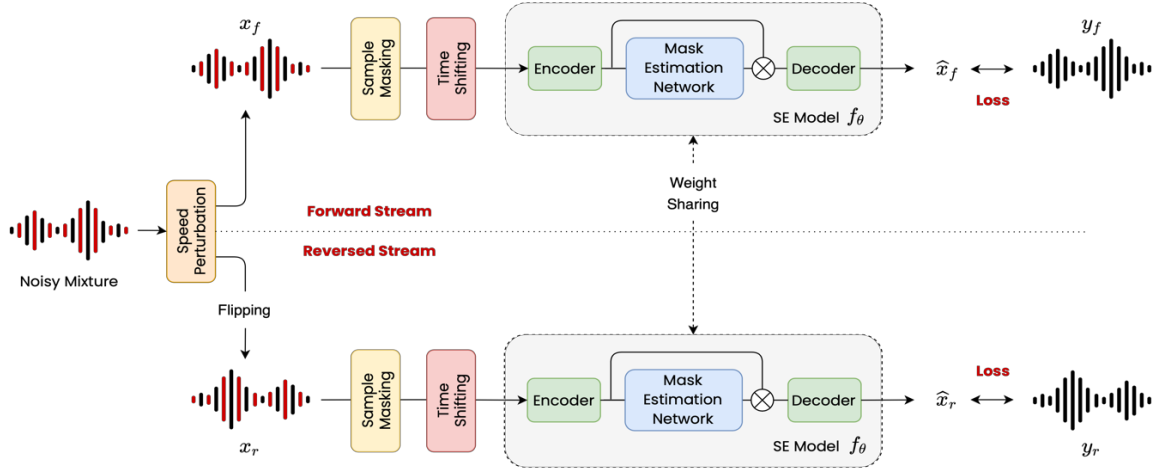

**Fig 1**. A schematic diagram of the TENET equipped with a masking-based SE model.

The remainder of this paper is organized as follows: Section 2 sheds light on the details of the integral components involved in TENET. Section 3 presents the experimental setup, results and discussion. Finally, we give the conclusion and envisage future research directions in Section 4.

## 2. TENET

As depicted in Figure 1, we introduce a novel SE framework, TENET, which consists of a time-reversal siamese network and a set of data augmentation techniques working in tandem, and it can be used in conjunction with most current mainstream SE models. In the following, we describe the ingredient components of TENET.

### 2.1. Time Reversal Siamese Network

First proposed for signature verification [13], the siamese network is a modeling technique that passes two different inputs through two weight sharing networks individually for discriminative learning. Lately, this technique has been widely used for both speech [15] and visual [16][17] representation learning. In this study, we adopt the siamese network and incorporate a time reversal method to deal with the SE task, which is called time reversal siamese network.

Since SE manages to remove the noise from input speech, we assume that the content of embedded clean speech is irrelevant to the noise removal for an SE task. Built on this concept, a time-reversal method can serve as a data augmentation technique, which flips the input noisy signal itself to augment the speech training data for SE. In the past literature, there has been associated research on the characteristics of the reversed speech. In [13], the author found that if we divide an utterance into segments of fixed duration, and all segments are then locally time-reversed without smoothing, the intelligibility of speech can remain unchanged for segments duration up to 50ms. In our preliminary experiments, frame-wise flipping utterances were not suitable for training an SE system and would lead to a poor convergence. Instead, we flip the complete waveform and employ the siamese network with the original and flipped utterances as the input into this work. As the schematic diagram shown in Figure 1, TENET contains a two-stream architecture with siamese SE models. The forward stream receives the noisy signal $x_f[n]$ which is converted to approximate its clean counterpart $y_f[n]$ as normal, while the reversed stream processes the reversed version of $x_f[n]$, viz. $x_r[n] = x_f[L-n]$, to approximate its clean duplicate $y_r[n] = y[L-n]$, where $L$ is the length of $x_f[n]$. Once the training is done, we can simply use one of the streams for inference. To the best of our knowledge, this is the first attempt to utilize time-reversed speech to train an SE model, and the design of the siamese architecture also allows flexibility for other alternatives of the training data or SE model structures.

The possible advantage of using time-reversed speech $x_r[n]$ together with the original (forward) speech $x_f[n]$ to train an SE model is twofold: First, $x_r[n]$ serves to increase the amount and variety of training data in order to realize data augmentation that has been shown to benefit model generalizability. Second, the spectrograms of $x_f[n]$ and $x_r[n]$ share the same auto-correlation function, revealing that they possess common temporal characteristics and can be additive to each other in learning of an SE model. This can be shown as follows. Let $X_{f,i}(e^{j\omega})$ and $X_{r,i}(e^{j\omega})$ denote the short-time Fourier transform of the $i^{th}$ frame for $x_f[n]$ and $x_r[n]$, respectively. We have

$$X_{r,i}(e^{j\omega}) = FT\{x_{r,i}[n]\}, \\ = FT\{x_{f,N-i}[M-n]\} = X^*_{f,N-i}(e^{j\omega})e^{-j\omega M}, \quad (1)$$

where $FT(\cdot)$ is the Fourier transform, $M$ is the frame size (assuming that there is no zero-padding in the last frame of $x_f[n]$), and "*" denotes the complex conjugate. Then the auto-correlation sequences of $X_{r,i}(e^{j\omega})$ and $X_{f,i}(e^{j\omega})$ along the $i$-axis, denoted by $R_r[\tau,\omega]$ and $R_f[\tau,\omega]$, can be shown equal according to Eq. (1):

$$R_r[\tau,\omega] = E\{X_{r,i}(e^{j\omega})X^*_{r,i+\tau}(e^{j\omega})\}, \\ = E\{X^*_{f,N-i}(e^{j\omega})e^{-j\omega M}X_{f,N-i-\tau}(e^{j\omega})e^{j\omega M}\}, \quad (2) \\ = R_f[\tau,\omega],$$

where $E\{\cdot\}$ is the expectation operator.

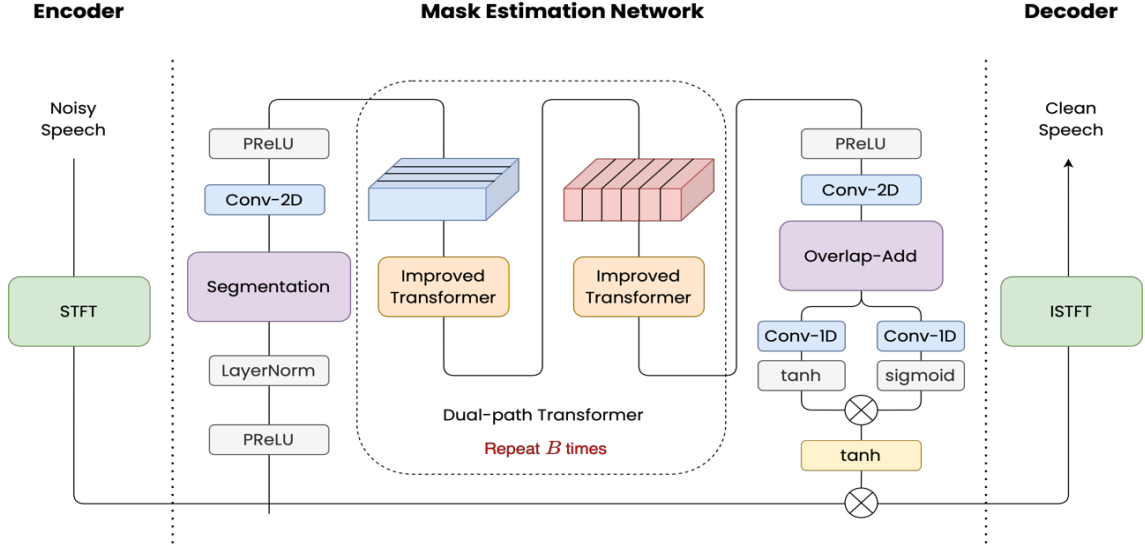

**Fig 2**. A schematic diagram of the complex dual-path transformer.

Basically, the presented TENET can allow any current mainstream SE model to process the two input streams and then share model parameters. In this work, we exploit complex dual-path transformer (CDPT), an effective masking-based SE model, for each stream of the siamese network in TENET. As shown in Figure 2, CDPT is composed of an encoder-decoder architecture and a mask estimation network, and we discuss its details in the following sub-sections.

### 2.2. Complex Dual-path Transformer

*2.2.1 Encoder-decoder*

As in [18], the author proposed to use the 1-D convolution and 1-D transpose convolution as the encoder-decoder in time domain, where the time-domain features extracted by 1-D convolution is called the "wavegrams" in [19]. Distinct from [18], CDPT replaces the wavegrams with the complex-valued spectrograms of the processed signal via conventional short-time Fourier transform (STFT) and inverse short-time Fourier transform (ISTFT), which can be written as

$$\mathbf{W} = \mathbf{U}\mathbf{X}, \quad (3)$$

$$\widehat{\mathbf{X}} = \mathbf{V}\mathbf{W}, \quad (4)$$

where $\mathbf{X} \in \mathbb{R}^{L \times T}$ is the matrix obtained by dividing the input waveform $x$ into $T$ overlapped segments of length $L$, and $\mathbf{W}, \mathbf{U}, \mathbf{V}$ are real-valued matrices denoting the complex spectrograms and discrete Fourier basis functions used in STFT and ISTFT, respectively. Note here that, the left half and right half of $\mathbf{W}$ are respectively arranged to be the real and imaginary parts of the spectrogram, and on top of that, we do not apply any non-linear operation to $\mathbf{W}$, so that the elements in $\mathbf{W}$ might be negative values.

*2.2.2 Mask Estimation Network*

To effectively consider the local and global dependency of frames, state-of-the-art speech separation models use attention-based dual-path modeling network for the mask estimation [20]. In this study, we adopt this idea to our mask estimation network, and the whole process consists of three stages: *segmentation*, *dual-path modeling*, and *overlap-add*.

- *Segmentation*

In the segmentation stage, for input features $\mathbf{W} \in \mathbb{R}^{N \times K}$ where $N$ is the feature dimension and $K$ is the number of time steps, we split $\mathbf{W}$ into $S$ chunks of a fixed-length $Q$ with a hop size $P = Q/2$, and then concatenate all chunks together to obtain a 3-D tensor $\mathbf{T} \in \mathbb{R}^{N \times Q \times S}$.

- *Dual-path Modeling*

The 3-D tensor $\mathbf{T}$ is then passed to the stack of $B$ dual-path modeling blocks, and each block contains the intra-chunk and inter-chunk processing. First, we denote the input tensor for each block as $\mathbf{T}_b \in \mathbb{R}^{N \times Q \times S}$, where $b$ is the block index, and $\mathbf{T}_1 = \mathbf{T}$. The intra-chunk processing first models the local information, which is applied to the second dimension of $\mathbf{T}_b$:

$$\mathbf{U}_b = [\, f_b^{intra}(\mathbf{T}_b[\,:,:,i\,]), i = 1, \ldots, S\,], \quad (5)$$

where $f_b^{intra}$ is the improved transformer defined by [20], and $\mathbf{T}_b[\,:,:,i\,]$ denotes the $i$-th chunk in $\mathbf{T}_b$.

$\mathbf{U}_b$ then serves as the input for inter-chunk processing, which models the global dependency throughout the chunks by applying the improved transformer to the last dimension of $\mathbf{U}_b$:

$$\mathbf{V}_b = [\, f_b^{inter}(\mathbf{U}_b[\,:,i,:\,]), i = 1, \ldots, Q\,], \quad (6)$$

where $f_b^{inter}$ is another improved transformer, and $\mathbf{U}_b[\,:,i,:\,]$ denotes the $i$-th time step in all $S$ chunks of $\mathbf{U}_b$.

- *Overlap-add*

After the dual-path modeling is completed, the overlap-add method is in turn applied to the $S$ chunks of $\mathbf{T}_{B+1}$ to generate the mask $\mathbf{M} \in \mathbb{R}^{L \times T}$. After that, we use the hyperbolic tangent function $tanh$ as a non-linear function $\varphi(\cdot)$ to limit the mask values within (-1, 1), and the enhanced features $\widehat{\mathbf{W}}$ is obtained by element-wise multiplication between $\varphi(\mathbf{M})$ and $\mathbf{W}$:

$$\widehat{\mathbf{W}} = \varphi(\mathbf{M}) \cdot \mathbf{W}. \tag{7}$$

The finally enhanced waveform can be decoded from $\widehat{\mathbf{W}}$ as Eq. (4).

### 2.3. Extra Data Augmentation

In representation learning, there have been studies [15][16][17] showing that increasing the variety of training data can learn the more powerful features through the siamese network. Thus, we adopt three data augmentation schemes: *speed perturbation*, *time shifting* and *sample masking*, respectively for TENET. Notably, these methods are all conducted on-the-fly during the network training in the waveform domain.

#### 2.3.1. Speed Perturbation

As for data augmentation of raw waveforms, speed perturbation [21] is a straightforward technique and widely deployed for acoustic modeling. Following the procedure of [21], we adopt this technique for SE and use the `speed` function of the `Sox` audio manipulation tool [22] to perturb the input signal by a factor $f_{sp}$, which is uniformly selected from 0.95 to 1.05 in this work.

#### 2.3.2. Time Shifting

Time shifting is a simple data augmentation method for audio, and it shifts the audio samples to the left/right with $f_{shift}$ second. In time shifting, we always apply the right shift to the audio, and $f_{shift}$ is randomly sampled from 0 to 0.625 seconds.

#### 2.3.3. Sample Masking

Mask prediction (MP) is the mechanism that masks parts of the input for prediction and commonly used for self-supervised training. For example, BERT [23] proposed to use the mask language model (MLM) as the pre-training objective. Furthermore, this method has also been proven effective for data augmentation in ASR [24]. Accordingly, here we employ sample masking as another data augmentation. Unlike [24], which performs masking on the log-mel spectrogram, here sample masking straightly masks $t$ consecutive audio samples $[t_0, t_0 + t - 1]$ to zero (making them silent) and we expect the model to predict the noise-free target waveform by considering the context information. In addition, there are two hyperparameters in sample masking, $t$ and $m_s$, which correspond to the mask length and the maximum number of masks, respectively. In our experiments, $t$ is fixed to 10 and $m_s$ is chosen from a uniform distribution in the interval [0, 150].

### 2.4. Training Objective

It is known that denoising algorithms in SE usually introduce artifacts in the processed speech. Although the artifact distortion does not have much impact on the human perception, it degrades the downstream ASR performance. In order to reduce the artifact distortion as well as maximize the perceptual quality of the enhanced speech, we propose to use a hybrid loss to training our SE system. To minimize the distortion between the clean speech $\mathbf{s}$ and enhanced speech $\hat{\mathbf{s}}$, we use the negative scale-invariant signal-to-distortion ratio (SI-SDR) [25] as a loss function, which can be formulated as follows:

$$\mathbf{s}_{target} = \frac{\langle \hat{\mathbf{s}}, \mathbf{s} \rangle \mathbf{s}}{\|\mathbf{s}\|^2}, \tag{8}$$

$$\mathbf{e}_{noise} = \hat{\mathbf{s}} - \mathbf{s}_{target}, \tag{9}$$

$$\mathcal{L}_{\text{SI-SDR}}(\mathbf{s}, \hat{\mathbf{s}}) = -10 \cdot \log_{10}\left(\frac{\|\mathbf{s}_{target}\|^2}{\|\mathbf{e}_{noise}\|^2}\right). \tag{10}$$

As for the perceptual quality optimization, we adopt phone-fortified perceptual (PFP) [9] loss to minimize the perceptual distance:

$$\mathcal{L}_{\text{PFP}}(\mathbf{s}, \hat{\mathbf{s}}) = \mathbb{E}_{\mathbf{s}, \hat{\mathbf{s}} \sim \mathcal{D}}[\|\Phi_{wav2vec}(\hat{\mathbf{s}}) - \Phi_{wav2vec}(\mathbf{s})\|_1], \tag{11}$$

where the $\mathbf{s}$ and $\hat{\mathbf{s}}$ are sampled from the training dataset $\mathcal{D}$ and $\Phi_{wav2vec}$ is the pretrained *wav2vec* encoder used to extract the 512-dimensional feature vectors for calculating the $L^1$ distance between $\mathbf{s}$ and $\hat{\mathbf{s}}$ in latent spaces. Following this, the hybrid loss function can be written as the following equation:

$$\mathcal{L}_{\text{hybrid}}(\mathbf{s}, \hat{\mathbf{s}}) = \mathcal{L}_{\text{SI-SDR}}(\mathbf{s}, \hat{\mathbf{s}}) + \alpha \cdot \mathcal{L}_{\text{PFP}}(\mathbf{s}, \hat{\mathbf{s}}), \tag{12}$$

where $\alpha$ is a weight parameter. Finally, we use a weighted sum of the hybrid losses for the reversed stream and the forward stream to train the TENET framework:

$$\mathcal{L}_{total}(\mathbf{s}, \hat{\mathbf{s}}) = \beta \cdot \mathcal{L}_{\text{hybrid}}^{forward}(\mathbf{s}, \hat{\mathbf{s}}) + \gamma \cdot \mathcal{L}_{\text{hybrid}}^{reversed}(\mathbf{s}, \hat{\mathbf{s}}), \tag{13}$$

where $\beta$ and $\gamma$ are weight parameters.

## 3. EXPERIMENTS

### 3.1. Experimental Setup

#### 3.1.1. Dataset

For evaluating the proposed methods, we conducted a series of experiments on the VoiceBank-DEMAND [26] benchmark dataset. In the training set, 11,572 utterances (from 28 speakers) are corrupted by 10 types of noise from the DEMAND [28] database at 4 different SNRs: 0, 5, 10 and 15 dB while the 824 utterances (from 2 speakers) are used for testing and contaminated by 5 types of noise at SNRs: 2.5, 7.5, 12.5 and 17.5 dB. For SE experiments, we set aside around 200 utterances from the training set for validation, and the others were for training. As for ASR systems, we used Kaldi [29] to build two hybrid DNN-HMM acoustic models, both of which were time-delay neural networks trained with lattice-free MMI (LF-MMI) objective function. The baseline DNN-HMM system (denoted by CCT-AM) was trained in the clean condition mode that used all of the clean utterances in VoiceBank [27] except the testing ones. In addition to CCT-AM, another DNN-HMM system (denoted by MCT-AM) was trained with noise-corrupted multi-condition data, which were synthesized by adding the clean recordings with 13 types of noise sources (distinct from the test set) from DEMAND with SNRs ranging from 0 to 15 dB.

Table 1: *PESQ, SI-SDR (dB) and WER (%) results of different SE systems on VoiceBank-DEMAND and VoiceBank-QUT-NOISE.*

| Acoustic Model | SE Model | Test set of VoiceBank | | | | | |
|---|---|---|---|---|---|---|---|
| | | DEMAND | | | QUT-NOISE | | |
| | | PESQ | SI-SDR (dB) | WER (%) | PESQ | SI-SDR (dB) | WER (%) |
| CCT-AM | No process | 1.97 | 8.45 | 23.76 | 1.25 | 3.88 | 82.32 |
| MCT-AM | No process | 1.97 | 8.45 | 8.31 | 1.25 | 3.88 | 38.87 |
| | DCCRN [31] | 2.77 | 18.94 | 7.31 | 1.77 | 11.54 | 27.56 |
| | PFPL [9] | 3.11 | 17.28 | 12.78 | 2.00 | 9.93 | 38.67 |
| | MetricGAN+ [12] | 3.15 | 8.52 | 8.32 | **2.28** | 2.83 | 43.63 |
| | TENET | **3.15** | **19.12** | **6.76** | 2.12 | **12.13** | **26.50** |

Table 2: *An ablation study of TENET on VoiceBank-DEMAND.*

| | PESQ | SI-SDR (dB) |
|---|---|---|
| TENET | **3.15** | 19.12 |
| - sample masking | 3.08 | **19.26** |
| - time shifting | 3.07 | 19.25 |
| - speed perturbation | 3.06 | 19.20 |
| - time reversal (CDPT only) | 3.01 | 19.06 |

To further examine the generalization capability of the presented methods, we created another test set by utilizing the same utterances as the original test set added with different types of noise to simulate the more severe noisy environments. To be specific, the noisy data was generated by mixing the noise sources complied from the QUT-NOISE [30] dataset, which contained non-stationary noise at -5, 0, 5 and 15 dB of SNRs, respectively. Note that, all of the waveforms used for the experiments were resampled to 16-kHz.

*3.1.2. Model Configurations*

As for the configuration of TENET, the DFT size of CDPT is set to 512, and the frame size and frame shift are set to 25ms and 6.25ms, respectively, with a Hanning window function. In the mask estimation network described in Subsection 2.2.2, the number of chunks $S$ is set to 100 with 1/2 overlapping between the neighboring chunks, parameter $B$ is set to 5 for iterative dual-path modeling, the number of the filters in the Conv-2D block of Figure 2 is set to 128, and the numbers of parallel heads and the hidden units in improved transformer are set to 8 and 256, individually.

In order to evaluate the proposed methods, we choose three state-of-the-art SE approaches for comparison: PFPL [9], DCCRN [31] and MetricGAN+ [12]. For PFPL and MetricGAN+, we use the pretrained models from repositories[1,2] for inference. On the other hand, DCCRN follows `DCCRN-CL` settings[3] and is trained to optimize the SI-SDR. For evaluation, we use PESQ (wide-band version) [6], SI-SDR [25], and word error rate (WER) as the speech quality, distortion, and intelligibility metrics to examine the performance of different SE systems, respectively. The corresponding results are summarized in Table 1.

**3.2. Comparison of state-of-the-art SE systems**

As presented in Table 1, we can easily observe that the further adoption of the multi-condition training data enables the resulting MCT-AM to obtain significant reductions in WER compared to CCT-AM. On top of MCT-AM, we investigate the performance of different SE methods and make some noteworthy observations as follows:

First, the two methods PFPL and MetricGAN+, which aim to optimize the perceptual quality, can lead to significant PESQ results in relation to DCCRN, whereas they perform worse on SI-SDR and WER metrics. One possible reason is that they introduce lots of artifacts in the processed speech, which in turn deteriorate the recognition accuracy. In contrast, although DCCRN performs much worse than other methods on speech quality recovery, it minimizes the distortion of the speech and thus delivers superior results on SI-SDR and WER over PFPL and MetricGAN+. Taking both factors into account, the proposed TENET method outperforms other methods and achieves state-of-the-art on the VoiceBank-DEMAND dataset in terms of PESQ, SI-SDR and WER.

Furthermore, we examine the generalization capability of TENET on the test set of scenarios contaminated with unseen noise sources (from the VoiceBank-QUT-Noise dataset). As we can see in the right half of Table 1, while MetricGAN+ achieves the highest PESQ score, it obtains the worst SI-SDR and WER among all methods and the unprocessed baseline. Comparatively, TENET can not only give high PESQ but also contribute to the optimal SI-SDR and the lowest WER. Hence, we further confirm the superiority of TENET.

**3.3. Ablation study**

In order to better realize the influence of every component in our proposed method, we conducted an ablation study for TENET. By observing the results presented in Table 2, we find that the *time-reversal* and *sample masking* procedures contribute the most on PESQ. While appending the sample masking component will slightly decrease the SI-SDR performance, combining all of the component procedures can promote the overall performance.

---
[1] https://github.com/aleXiehta/PhoneFortifiedPerceptualLoss
[2] https://huggingface.co/speechbrain/metricgan-plus-voicebank
[3] https://github.com/huyanxin/DeepComplexCRN

Table 3: *Equipping TENET with DCCRN.*

|            | PESQ | SI-SDR (dB) |
|------------|------|-------------|
| No process | 1.97 | 8.45        |
| DCCRN      | 2.77 | 18.94       |
| + TENET    | **2.97** | **19.00** |

### 3.4. Model flexibility and compatibility

To demonstrate the flexibility and compatibility of TENET, we conduct an extra experiment that equips TENET with another SE model, DCCRN, to see whether it can improve the performance as well. According to the results shown in Table 3, when adopting the TENET framework, DCCRN behaves even better for the SE task in terms of PESQ and SI-SDR metrics.

### 4. CONCLUSION AND FUTURE WORK

In this study, we have proposed a novel speech enhancement framework, TENET, which incorporates the time-reversal method and siamese network to form a robust SE architecture for noise-robust ASR. Experimental results on the VoiceBank-DEMAND benchmark dataset have demonstrated that TENET can achieve the state-of-the-art results in terms of both SE and ASR metrics, also confirming the feasibility and versatility of TENET. We thus believe that TENET can provide a new avenue for the future SE developments, and the time-reversal and sample masking methods can be further applied to other SE and ASR tasks.